\documentclass{ws-ijmpe}
\usepackage[super,compress]{cite}
\begin{document}

\markboth{Raghunath Sahoo}{Heavy Flavor Production at the Large Hadron Collider: A Machine Learning Approach}

\catchline{}{}{}{}{}
%%%%%%%%%%%%%%%%%%%%%%%%%%%%%%%%%%%%%%%%%%%%%%%%%%%%%%%%%%%%%%%%%%%%

\title{Heavy Flavor Production at the Large Hadron Collider: A Machine Learning Approach}

\author{Raghunath Sahoo\footnote{Email: Raghunath.Sahoo@cern.ch, Invited Plenary in ``Hot QCD Matter-2024"}}

\address{Department of Physics, Indian Institute of Technology Indore, Simrol, Khandwa Road, Indore-453552, Madhya Pradesh, INDIA}

%University Department, University Name, Address\\
%City, State ZIP/Zone,
%Country\footnote{State completely without abbreviations, the
%affiliation and mailing address, including country. Typeset in 8~pt
%Times italic.}\\
%first\_author@university.edu}

\maketitle

\begin{history}
\received{Day Month Year}
\revised{Day Month Year}
%\accepted{Day Month Year}
%\comby{(xxxxxxxxxx)}
\end{history}

\begin{abstract}
Charmonia suppression has been considered as a smoking gun signature of quark-gluon plasma. However,
the Large Hadron Collider has observed a lower degree of suppression as compared to the Relativistic Heavy Ion Collider energies, due to regeneration effects in heavy-ion collisions. Though proton collisions are considered 
to be the baseline measurements to characterize a hot and dense medium formation in heavy-ion collisions,
LHC proton collisions with its new physics of heavy-ion-like QGP signatures have created new challenges.
To understand this, the inclusive charmonia production at the forward rapidities in the dimuon channel is compared with the corresponding measurements in the dielectron channel at the midrapidity as a function of final state 
charged particle multiplicity. None of the theoretical models quantitatively reproduce the
experimental findings leaving out a lot of room for theory. To circumvent this and find a reasonable understanding,
we use machine learning tools to separate prompt and nonprompt charmonia and open charm mesons using the
decay daughter track properties and the decay topologies of the mother particles. Using PYTHIA8 data, we 
train the machine learning models and successfully separate prompt and nonprompt charm hadrons from the
inclusive sample to study various directions of their production dynamics. This study enables a domain of using
machine learning techniques, which can be used in the experimental analysis to better understand 
charm hadron production and build possible theoretical understanding.

\end{abstract}


\ccode{PACS numbers:}

%\tableofcontents

\section{Introduction}
In the heavy flavor sector, charmonia (J/$\psi$), the bound state of a charm and anti-charm quark 
($\rm c\bar{c}$) plays an
important role, the suppression of which in heavy-ion collisions at ultra-relativistic energies is considered a 
signature of the deconfined primordial matter, called quark-gluon plasma (QGP). This, on the other hand, leads 
to the enhancement of open charm mesons like $D^0$. A complementary study taking both J/$\psi$ and $D^0$
can give a full picture of probing the produced QCD medium using charmonia. While making such a study in
heavy-ion collisions to probe the hot QCD medium, usually the minimum-bias proton-proton (pp) collisions are 
taken as a baseline measurement. It should also be noted here that the degree of charmonia suppression at the LHC energies is observed to be lower compared to RHIC energies, because of the availability of higher energy
phase space at the LHC leading to charmonia regeneration effects\cite{ALICE:2015jrl}.  At the TeV LHC energies, several partonic collisions may occur
in a single pp collision, which affects the total multiplicity through the production of light quarks and gluons.
These high-multiplicity pp events are of special importance at the LHC energies, where one observes
several heavy-ion-like signatures, which include collective flow pattern\cite{CMS:2010ifv}, enhancement of strangeness\cite{strange,strange1} etc. For a general review of such QGP-like signatures in proton collisions, please see Ref.\cite{Sahoo:2021aoy}. On the contrary, the absence of evidence of jet quenching in such
high-multiplicity pp events makes them illusive. These observations add to the earlier conjecture that high-energy 
pp collisions could produce statistical systems capable of showing hydrodynamics behavior\cite{Fermi,Hagedorn,Landau}. Such statistical
systems with partonic quanta, which are locally thermalized and behave like strongly interacting hot fluid of QGP
are to be confronted with experimental tests. However, high-multiplicity proton-antiproton collisions at the 
Tevatron energies ($\sim \sqrt{s} = $ 1.8 TeV) didn't support the formation of QGP in such collisions\cite{Tevatron}.

In this report, we present the forward rapidity inclusive J/$\psi$ production as a function of the final state charged
particle multiplicity measured with ALICE\cite{ALICE:2021zkd}, which is not quantitatively explained by the existing theoretical models. Further, we employ Machine Learning (ML) techniques to separate prompt and nonprompt J/$\psi$ and
$D^0$ meson in pp collisions at TeV energies to study various production dynamics of charmonia and open
charms using PYTHIA8 event generator using track-level properties and
decay topology\cite{Prasad:2023zdd,Goswami:2024xrx}.

\subsection{Charmonia measurement in proton collisions}

\begin{figure*}[ht!]
\begin{center}
\includegraphics[scale = 0.35]{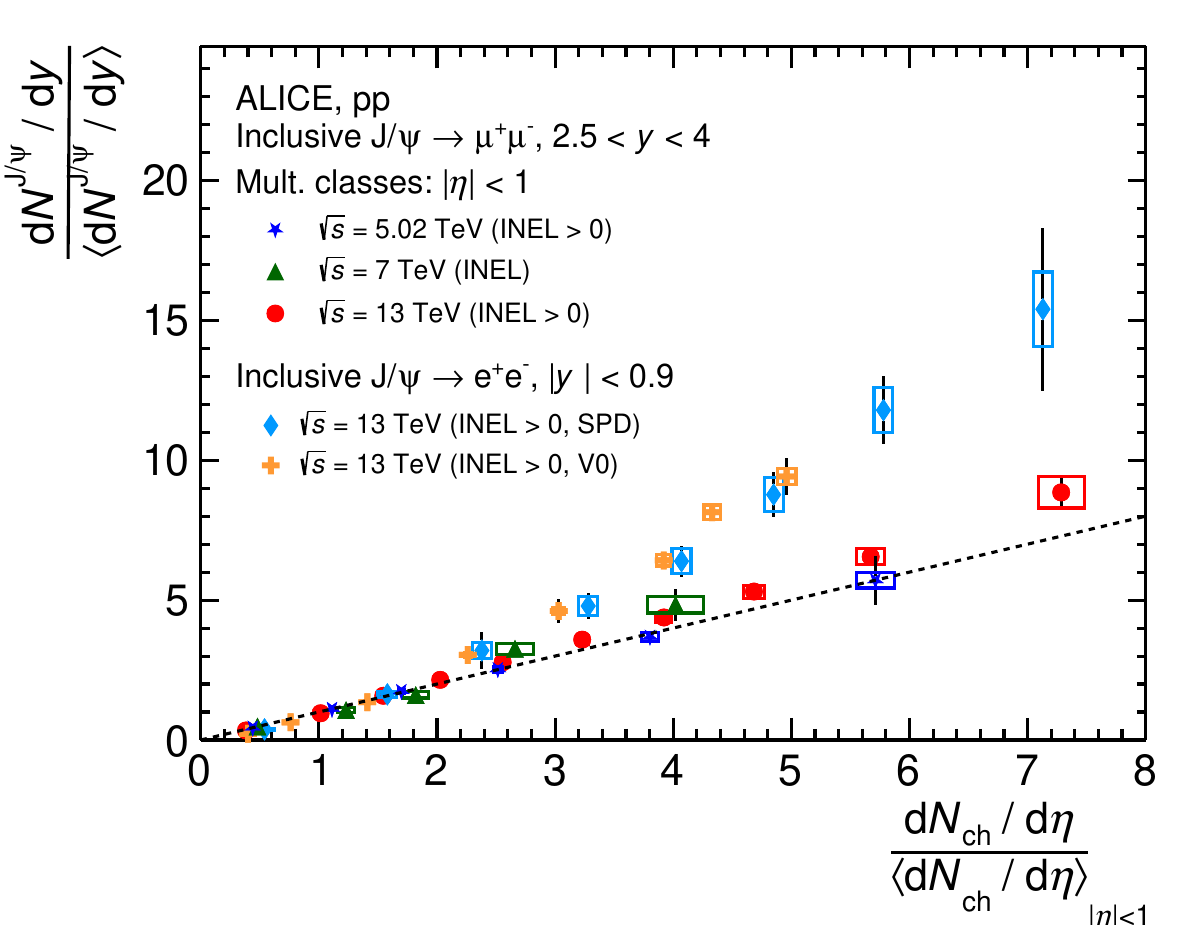}
\caption{(Colour Online) Forward rapidity relative J/$\psi$ yields in pp collisions at $\sqrt{s}$ = 5.02, 7 and 13 TeV compared to $\sqrt{s}$ = 13 TeV measurement at midrapidity \cite{ALICE:2021zkd}.}
\label{fig1}
\end{center}
\end{figure*}

\begin{figure*}[ht!]
\begin{center}
\includegraphics[scale = 0.43]{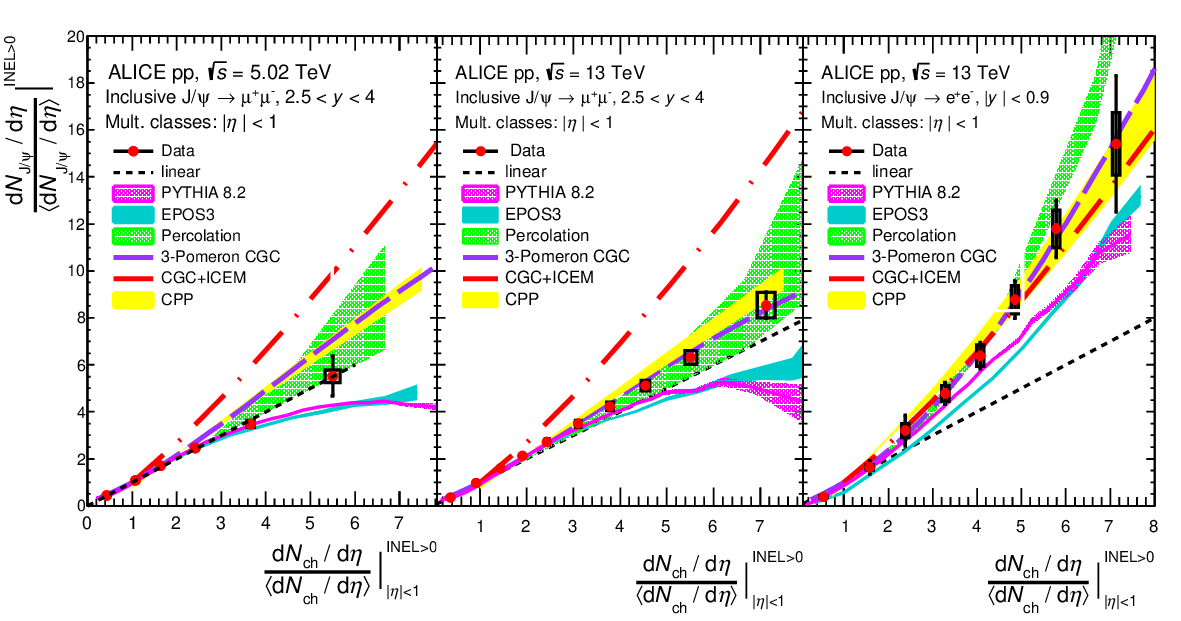}
\caption{(Colour Online) Forward rapidity relative J/$\psi$ yields in pp collisions at $\sqrt{s}$ = 5.02 and 13 TeV compared various theoretical model predictions as discussed in the text\cite{ALICE:2021zkd}.}
\label{fig2}
\end{center}
\end{figure*}

ALICE has measured the forward rapidity $(2.5 < y < 4)$ J/$\psi$ in pp collisions at $\sqrt {s} = 5.02, 7$ and 13 TeV
in the dimuon channel using a muon spectrometer. Whereas the proxy of centrality (or impact parameter) is the
final state charged particle multiplicity, which is measured at the midrapidity ($|\eta| < 1$) using a Silicon Pixel Detector (SPD) to avoid possible
autocorrelation bias in the measurement. The details of the measurement procedure can be found in Ref.\cite{ALICE:2021zkd}.

The self-normalized yield of J/$\psi$ is defined as the ratio of the yield in a given multiplicity window to the average yield across all the measured multiplicity bins, i.e. $(dN_{J/\psi}/dy)$/$<dN_{J/\psi}/dy>$.  This is shown in Fig.\ref{fig1} as a function of the corresponding self-normalized charged particle yield. For all the discussed energies 
the relative J/$\psi$ yield shows a nearly linear rise with the midrapidity charged particle relative multiplicity. 
A similarity in the forward rapidity J/$\psi$ production across various energies suggests in a given multiplicity window, J/$\psi$ 
production is more or less independent of collision energy. These forward rapidity measurements are compared with
the corresponding measurements at midrapidity in the dielectron channel for pp collisions at $\sqrt{s}$ = 13 TeV, where the multiplicity estimators are taken both at midrapidity and forward rapidity, $-3.7 < \eta < -1.7 $ and $2.8 < \eta < 5.1$  (V0 detector). Although the results are consistent with experimental uncertainties showing the absence
of possible auto-correlation bias arising from multiplicity selection bias, there is a deviation from the linear behavior
of J/$\psi$ production in the midrapidity dielectron channel as compared with the forward rapidity dimuon channel. 
To have a better understanding of J/$\psi$ production, these results are compared with available theoretical models
like Coherent Particle Production (CPP), CGC with ICEM (improved color evaporation model), 3-Pomeron CGC,
Percolation, EPOS3, and PYTHIA8\cite{ALICE:2021zkd}. Both EPOS and PYTHIA8 describe the forward rapidity
J/$\psi$ yield at low-multiplicity while underestimating the high-multiplicity data. The CPP model with a phenomenological parametrization for mean multiplicities of light hadrons and J/$\psi$, shows a very good agreement with the high-multiplicity measurements both for $\sqrt{s}$ = 5.02 and 13 TeV. CGC+ICEM employs the
NRQCD framework to describe J/$\psi$ hadronization. While this model describes the midrapidity dielectron
channel results, it predicts a faster-than-linear increase of J/$\psi$ yield with multiplicity for pp collisions. The 3-gluon fusion model seems to describe the multiplicity-dependent J/$\psi$ yield both at midrapidity and forward rapidity
for $\sqrt{s}$ = 13 TeV, whereas it fails to describe the high-multiplicity region for $\sqrt{s}$ = 5.02 TeV. Percolation
model with larger uncertainties however seems to describe the forward rapidity multiplicity-dependent J/$\psi$
production both for $\sqrt{s}$ = 5.02 and 13 TeV. None of the theoretical models seem to describe the multiplicity-dependent inclusive J/$\psi$ production in pp collisions across rapidity, multiplicity, and production channels. Let us
now move to use ML techniques to separate prompt and nonprompt  J/$\psi$  and $D^0$ for a better understanding
of the underlying production mechanisms.
 
\subsection{Machine Learning tools heavy flavor studies}

\begin{figure*}[ht!]
\begin{center}
\includegraphics[scale = 0.42]{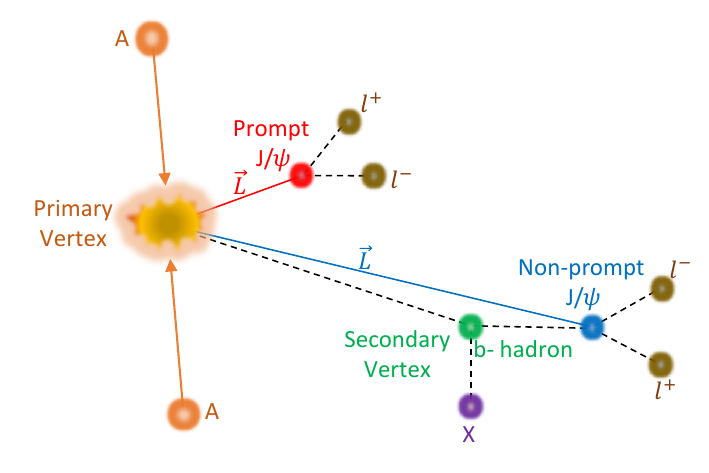}
\includegraphics[scale = 0.35]{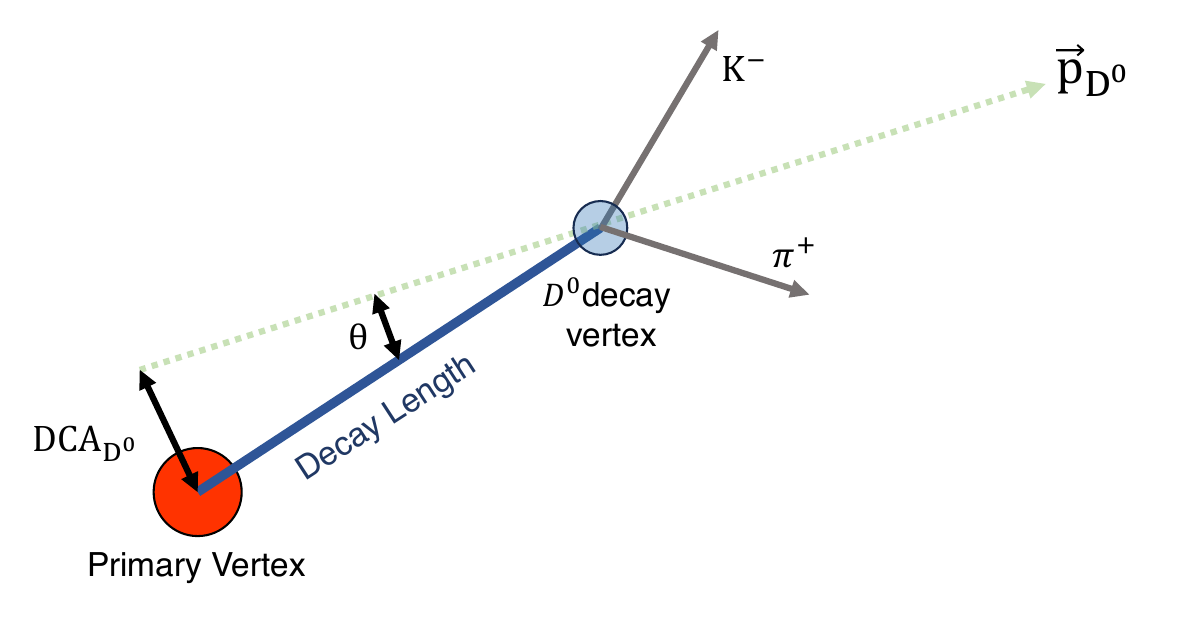}
\caption{(Colour Online) (Left) Decay topology of prompt and nonprompt J/$\psi$\cite{Prasad:2023zdd}, and (Right) for $D^0$ \cite{Goswami:2024xrx}.}
\label{fig3}
\end{center}
\end{figure*}

\begin{figure*}[ht!]
\begin{center}
\includegraphics[scale = 0.15]{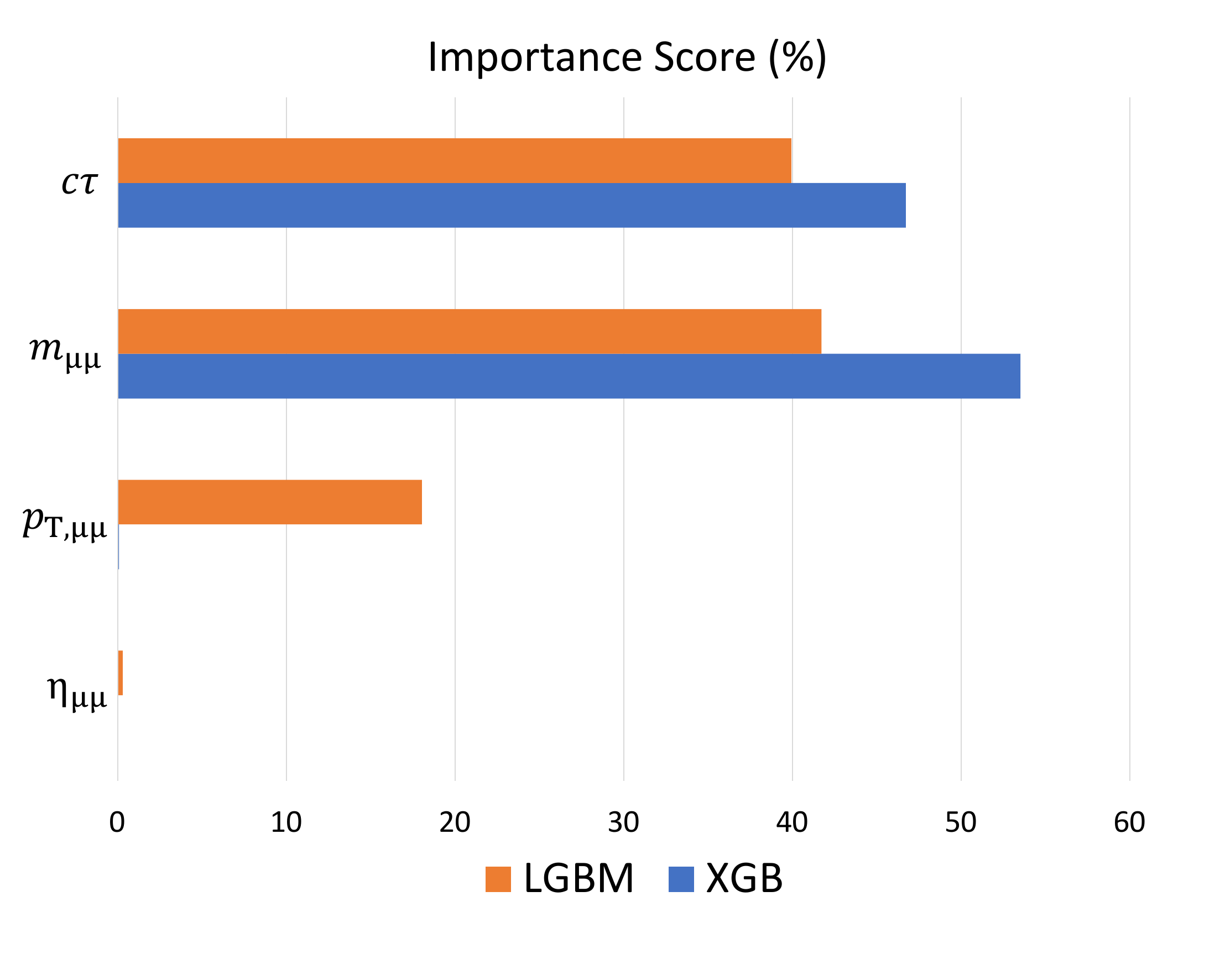}
\includegraphics[scale = 0.2]{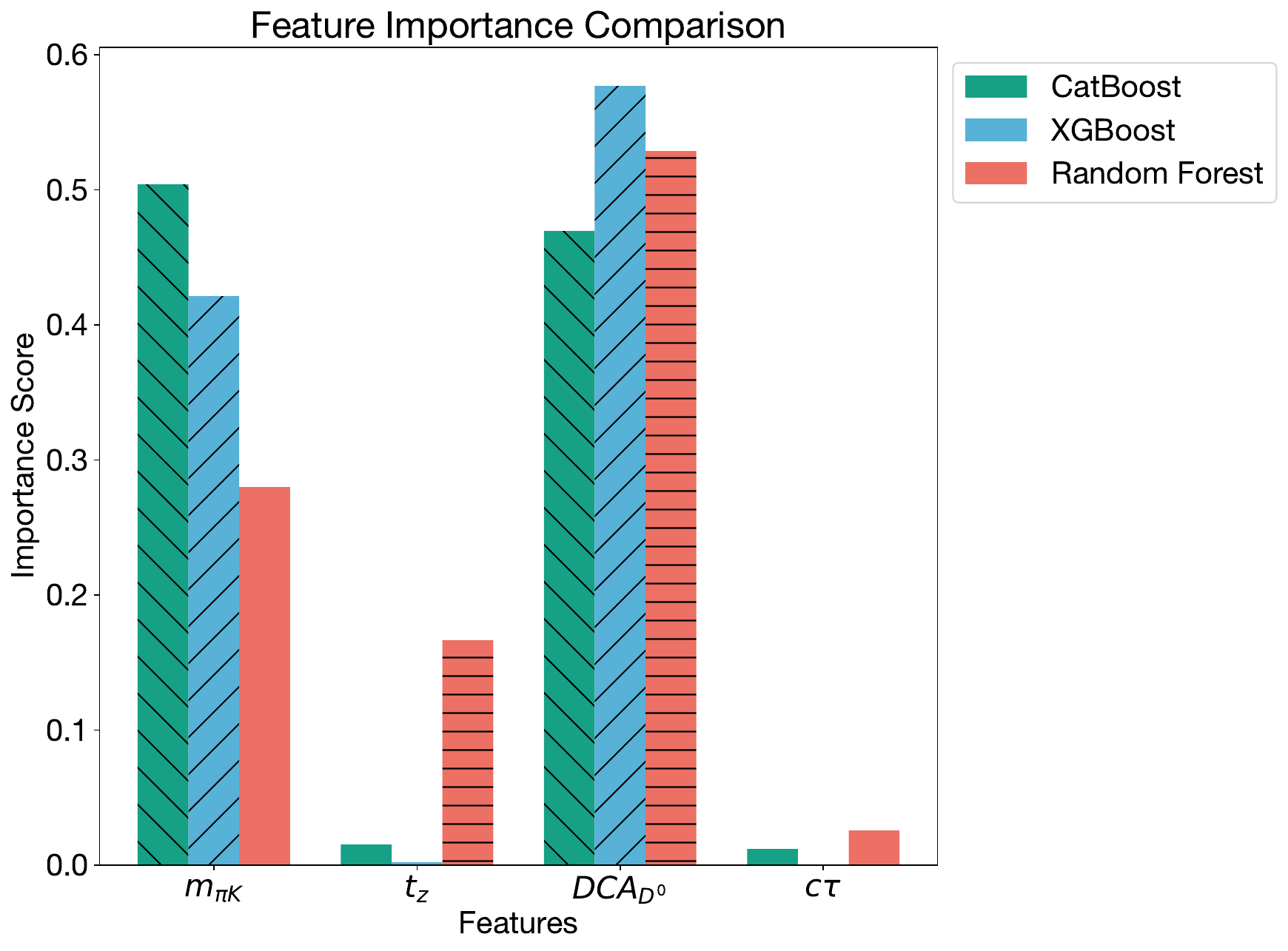}
\caption{(Colour Online) (Left) Comparison of importance scores (\%) for the input variables used for prompt and nonprompt J/$\psi$ separation\cite{Prasad:2023zdd}, and (Right) for separating prompt and nonprompt $D^0$ \cite{Goswami:2024xrx}.}
\label{fig4}
\end{center}
\end{figure*}

The inclusive production of charmonia consists of J/$\psi$ produced directly in the hadronic/nuclear collisions and
those produced via the feed down from directly produced higher charmonium states like $\psi(2S)$ and $\chi_c$.
These are called prompt J/$\psi$. Whereas those produced from the weak decay of beauty hadrons are classified as nonprompt J/$\psi$. The prompt J/$\psi$ is produced close to the interaction vertex, whereas the nonprompt
J/$\psi$ decay is associated with a secondary vertex, as shown in the decay topology (Fig.\ref{fig3}).  This classification can help
in the indirect estimation of the nuclear modification factor in the beauty sector in addition to the spin polarization 
measurements\cite{Prasad:2023zdd}. In the experimental data analysis, this separation method relies on statistical
methods with template fitting. However, the discussed technique uses track-level properties of the daughter
particles with the decay topology of  J/$\psi$.

We use ML models that use gradient-boosting-decision-trees-based classifications like XGBoost and LightGBM with simulated data of properly tuned PYTHIA8 for pp collisions at $\sqrt{s}$ = 13 TeV for training and prediction. Details of the methodology can be found in Ref.\cite{Prasad:2023zdd}. We estimate the pseudoproper decay length ($c\tau$), dimuon invariant mass, transverse momentum, and pseudorapidity as the
input variables for the topological separation of J/$\psi$. The $D^{0}$ meson uses the invariant mass, pseudo-proper time ($t_{z}$), distance of closest approach and $c\tau$. Model parameters such as the loss function, learning rate, number of trees, and maximum depth are 
tuned for each model. The optimized parameters are selected through a grid search method. As shown in Fig. \ref{fig4}, the importance score shows that both $c\tau$ and dimuon invariant mass play a significant role in the training and prediction compared to other parameters to identify prompt and nonprompt production of $\rm{J}/\psi$. Similarly, for the separation of prompt and nonprompt $D^0$, as shown in
Fig. \ref{fig4}, both the invariant mass and the distance of the closest approach are important parameters in the
ML techniques used.

\begin{figure*}[ht!]
\begin{center}
\includegraphics[scale = 0.3]{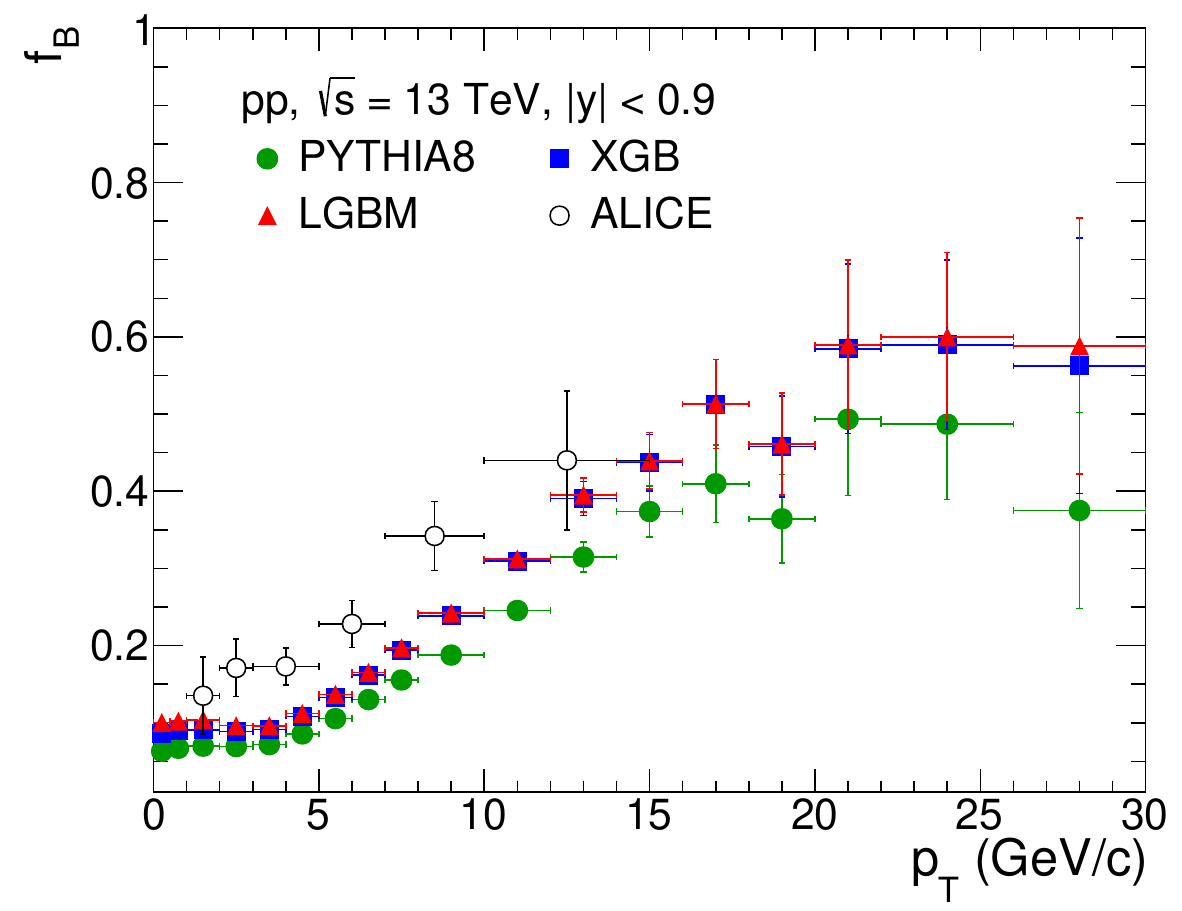}
\includegraphics[scale = 0.3]{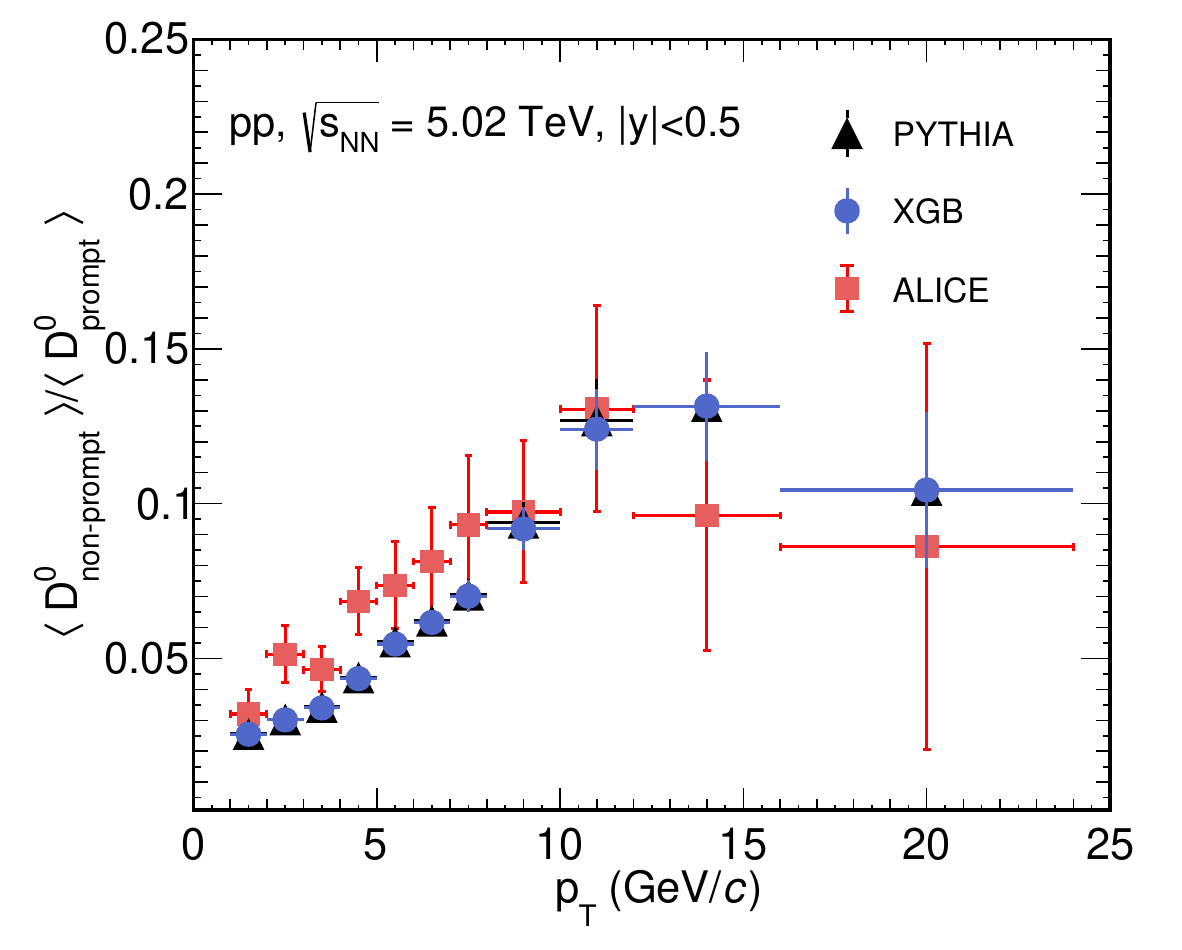}
\caption{(Colour Online) (Left) J/$\psi$ fraction from b-hadron decay ($f_B$) as a function of $p_{\rm T}$ at midrapidity for minimum-bias pp collisions at $\sqrt{s}$ = 13 TeV using PYTHIA8, predictions from XGB and LGBM
and a comparison with ALICE data is shown\cite{Prasad:2023zdd}. (Right) Nonprompt to prompt $D^0$ ratio for minimum-bias pp collisions at $\sqrt{s}$ = 5.02 TeV using PYTHIA8. Predictions from XGB and a comparison with ALICE data are also shown\cite{Goswami:2024xrx}.}
\label{fig5}
\end{center}
\end{figure*}

In Fig. \ref{fig5}, we show the J/$\psi$ fraction from b-hadron decay ($f_B$) as a function of $p_{\rm T}$ at midrapidity for minimum-bias pp collisions at $\sqrt{s}$ = 13 TeV using PYTHIA8, along with the predictions from XGB and LGBM,
and a comparison with ALICE data\cite{Prasad:2023zdd}. Here the trend of $f_B$ is similar to the experimental 
results and the ML models seem to do a good job even estimating $f_B$ in finer bins of $p_{\rm T}$ using ML.
Fig. \ref{fig5} (right) shows the ratio of nonprompt to prompt $D^0$ at midrapidity ($|y| < 0.5$) for minimumbias pp collisions at $\sqrt{s}$ = 5.02 TeV using PYTHIA8. Predictions from XGB and a comparison with ALICE data are also shown\cite{Goswami:2024xrx}. This ratio quantifies the fraction of $D^0$ coming from beauty hadron decays
in comparison to the direct production from the charm sector. There is a linear increase in this ratio with $p_{\rm T}$
upto $p_{\rm T} \simeq 12$ GeV/c. XGB seems to agree quantitatively with the estimates of
PYTHIA8. Details of the associated results for both charmonia and open charm meson can be found in the
Refs. \cite{Prasad:2023zdd,Goswami:2024xrx}. 

\section{Summary}
The experimental measurement of inclusive J/$\psi$ at forward rapidity in the dimuon channel is shown for TeV pp collisions as a function of final state charged particle multiplicity. These measurements are compared to those obtained in the dielectron channel. A linear trend is seen in the former case, whereas an increase
higher than linear is observed in the latter case. None of the discussed theoretical models seem to explain the data quantitatively in the discussed region of charged particle multiplicity. Further, with the need to separate prompt from
non-prompt charmonia and open charms, we use ML techniques, which are found to do an excellent job taking the 
track-level properties and decay topologies. This method, augmented to the mainstream of data analysis, will be a boon, given that experimental methods use statistical template fitting or additional detectors for secondary vertexing is a need for such separation. For a general review of charm and beauty as next-generation measurements to study QCD plasma created in TeV hadronic and nuclear collisions, please see Ref.\cite{Das:2021igk}.

\section*{Acknowledgements}
 The author gratefully acknowledges the DAE-DST, Government of India funding
under the mega-science project “Indian participation in
the ALICE experiment at CERN” bearing Project No. SR/MF/PS-02/2021-IITI(E-37123)
under which these research works have been carried out.

\end{document}